\documentclass[twocolumn,footinbib]{revtex4}

\IfFileExists{srcltx.sty}{\usepackage[active]{srcltx}}

\usepackage{amssymb,amsmath,bm}%
\usepackage{graphicx}%

         \newcount\hour \newcount\minute \hour=\time \divide \hour by 60
         \minute=\time \count99=\hour \multiply \count99 by -60 \advance
         \minute by \count99  

\usepackage{xspace,paralist} %

\usepackage[colorlinks=true,bookmarks=false]{hyperref}

\IfFileExists{hypernat.sty}{\usepackage{hypernat}}

\newcommand{\mev}{\ensuremath{\:\mathrm{MeV}}} %
\newcommand{\kev}{\ensuremath{\:\mathrm{keV}}} %
\newcommand{\mpc}{\:\mathrm{Mpc}} 
\newcommand{\numsm}{$\nu$MSM\xspace} 

\newcommand{\lcdm}{\ensuremath{\Lambda}\textsc{CDM}\xspace} %
\newcommand{\lcwdm}{{\ensuremath{\Lambda}\textsc{CWDM}}\xspace} %
\newcommand{\wdm}{\textsc{wdm}\xspace} %
\newcommand{\fnrp}{{\ensuremath{f_\textsc{nrp}}}\xspace}%
\newcommand{\Fwdm}{{\ensuremath{F_\wdm}}\xspace}%
\newcommand{\nrp}{\textsc{nrp}} %
\newcommand{\rp}{\textsc{rp}} %
\newcommand{\lya}{Ly-$\alpha$\xspace} %
\begin{document}

\title{\begin{flushright}
    \footnotesize \texttt{CERN-PH-TH/2008-234, LAPTH-1290/08}
  \end{flushright}Realistic sterile neutrino dark matter with keV mass\\
  does not contradict cosmological bounds}
\author{Alexey~Boyarsky$^{a,b}$, Julien Lesgourgues$^{c,d,e}$,
  Oleg~Ruchayskiy$^d$,
  Matteo Viel$^{f,g}$\\
  \small\it $^{a}$ETHZ, Z\"urich, CH-8093,  Switzerland\\
  \small\it   $^{b}$Bogolyubov Institute for Theoretical Physics, Kiev 03680, Ukraine\\
  \small\it $^c$PH-TH, CERN, CH-1211 Geneve 23, Switzerland\\
  \small\it $^{d}$\'Ecole Polytechnique F\'ed\'erale de Lausanne,
  \small\it FSB/ITP/LPPC, BSP, CH-1015, Lausanne, Switzerland\\
  \small\it $^e$ LAPTH, Universit\'e de Savoie, CNRS,
  B.P.110, F-74941 Annecy-le-Vieux Cedex, France\\
  \small\it $^f$INAF -- Osservatorio Astronomico di Trieste, Via G.B.~Tiepolo
  11,
  I-34131 Trieste, Italy\\
  \small\it $^g$INFN -- National Institute for Nuclear Physics, Via Valerio 2,
  I-34127 Trieste, Italy\\}

\date{\today}
 \begin{abstract}
   Previous fits of sterile neutrino dark matter models to
   cosmological data assumed a peculiar production mechanism, which is
   not representative of the best-motivated particle physics models
   given current data on neutrino oscillations. These analyses ruled
   out sterile neutrino masses smaller than 8-10~keV. Here we focus on
   sterile neutrinos produced resonantly. We show that their
   cosmological signature can be approximated by that of mixed Cold
   plus Warm Dark Matter (CWDM). We use recent results on
   $\Lambda$CWDM models to show that for each mass $\ge 2$~keV, there
   exists at least one model of sterile neutrino accounting for the
   totality of dark matter, and consistent with Lyman-$\alpha$  and
   other cosmological data. Resonant production occurs in the
   framework of the $\nu$MSM (the extension of the Standard Model with
   three right-handed neutrinos). The models we checked to be allowed
   correspond to parameter values consistent with neutrino oscillation
   data, baryogenesis and all other dark matter bounds.
 \end{abstract}

\maketitle

The sterile neutrino is a very interesting Dark Matter
candidate~\cite{Dodelson:93,Shi:98,Dolgov:00,Abazajian:01a,Abazajian:01b,Asaka:05a}.
The existence of sterile neutrinos (right-handed or gauge singlet) is one of
the most simple and natural explanations of the observed flavor oscillations
of active neutrinos. It was observed long ago that such particles can be
produced in the Early Universe through oscillations with active
neutrinos~\cite{Dodelson:93}.  For any mass (above $\sim 0.4$~keV, which is a
universal lower bound on any fermionic Dark Matter (DM) particle, see
\cite{Boyarsky:08a} and references therein) sterile neutrinos produced in this
way can end up with a correct relic
density~\cite{Dodelson:93,Shi:98,Abazajian:01a,Asaka:06c,%
  Shaposhnikov:08a,Laine:08a}.

A single right-handed neutrino would be unable to explain the two observed
mass splittings between Standard Model (SM) neutrinos. Moreover, should this
neutrino play the role of DM, its mixing with active neutrinos would be too
small for explaining the observed flavor
oscillations~\cite{Asaka:05a,Boyarsky:06a}.  However, in presence of three
right-handed neutrinos (one for each SM flavor), active neutrino mass
splittings and DM may be explained at the same time~\cite{Asaka:05a}.
Moreover, the mass of each sterile neutrino can be chosen below the
electroweak scale and additionally explain the matter-antimatter asymmetry
of the Universe (baryogenesis)~\cite{Asaka:05a}. These observations motivated
a lot of recent efforts for developing this model, called the
\numsm~\cite{Asaka:06c,Shaposhnikov:06b,Gorbunov:07a,Bezrukov:07,Shaposhnikov:08b,Shaposhnikov:08d},
and for constraining sterile neutrino
DM~\cite{Boyarsky:06c,Riemer:06,Boyarsky:06d}.

Because of its mixing with flavor neutrinos, this DM particle has a
small probability of decaying into an active neutrino and a photon of
energy $E= m_{s}/2$~\cite{Pal:81}, producing a monochromatic line in
the spectrum of DM dominated objects. The corresponding photons flux
depends on the sterile neutrino mass $m_{s}$ and mixing angle $\theta$
as $F \sim \theta^{2} m_{s}^{5} $. For each value of the mass and of
other parameters in the model (see below), the angle $\theta$ is fixed
by the requirement of a correct DM abundance. 
Combining this constraint with bounds on decay lines in astrophysical spectra
allows to put an \emph{upper} limit on the mass of DM sterile
neutrinos~\cite{Dolgov:00,Abazajian:01b,Boyarsky:06c,Riemer:06,Boyarsky:06d}.

Within the \numsm, the relation between $m_s$, $\theta$ and the DM
abundance can be affected by the presence of a \emph{lepton asymmetry}
(an excess of leptons over anti-leptons). In this case, the production
of sterile neutrinos may be of the resonant type~\cite{Shi:98}. The
lepton asymmetry required for this mechanism to be effective is
several orders of magnitude larger than the baryon asymmetry $\eta_B
\sim 10^{-10}$. In many models of baryogenesis (for a review see
e.g.~\cite{Davidson:2008bu}), both asymmetries are of the same order,
because they are generated above the electroweak scale and sphaleron
processes equalize them. Instead, in the \numsm, the lepton asymmetry
is generated below the electroweak scale, when sphaleron processes are
not active anymore~\cite{Shaposhnikov:08a}. As a result, it can be as
large as the upper limit imposed by Big Bang Nucleosynthesis (BBN) and
other cosmological constraints (see e.g.~\cite{Serpico:05} and
refs. therein). Such a large lepton asymmetry is consistent with
generic values of the parameters of the \numsm, satisfying current
data on neutrino oscillations, cosmological requirements
(baryogenesis, BBN constraints) and particle physics
constraints~\cite{Daum:2000ac}. So, resonant production (RP) is a
natural way of producing sterile neutrino DM in the \numsm. At the
same time, most previous constrains on sterile neutrino DM assumed
non-resonant production (NRP)~\cite{Dodelson:93,Asaka:06c}.


In the NRP case, the comparison of X-ray
bounds~\cite{Boyarsky:06c,Riemer:06,Boyarsky:06d} with constraints on DM relic
abundance~\cite{Asaka:06c} gives an \emph{upper bound} $m_{s}^\nrp \le 4~
\kev$ on the DM sterile neutrino mass. In the more effective RP scenario,
smaller mixing angles are required and the corresponding bound is much weaker:
$m_{s}^\rp \lesssim 50$~keV~\cite{Laine:08a}.

The most robust \emph{lower bound} on the DM mass comes from the analysis of
the phase space density of compact objects, e.g. dwarf spheroidals of the
Milky Way halo.  The universal Gunn-Tremaine bound~\cite{Tremaine:79} can be
made stronger if one assumes a particular primordial phase-space distribution
function.  For NRP sterile neutrinos, this leads to $m_s^{\nrp} > 1.8\kev$,
while for RP particles the bound is weaker: $m_s^{\rp} > 1\kev
$~\cite{Boyarsky:08a}.

An interesting property of sterile neutrino DM with keV mass is that it falls
in the Warm Dark Matter (WDM) category.  Lyman-$\alpha$ (\lya) forest
observations in quasar spectra provide strong \emph{lower bounds} on the mass
of DM sterile neutrinos produced with the NRP
mechanism~\cite{Viel:06,Seljak:06,Boyarsky:08c}.  The analysis of SDSS \lya
data led to $m_{s}^\nrp>13$~keV in Ref.~\cite{Seljak:06}, or
$m_{s}^\nrp>10$~keV in~\cite{Viel:06}. In \cite{Boyarsky:08c} these bounds
were revisited using the same SDSS \lya data (combined with
WMAP5~\cite{Dunkley:2008ie}), but paying special attention to the
interpretation of statistics in the parameter extraction, and to possible
systematic uncertainties. It was shown that a conservative (frequentist,
3-$\sigma$) lower bound is $m_{s}^\nrp>8$~keV. The \lya method is still under
development, and there is a possibility that some of the related physical
processes are not yet fully understood.  However, at this moment, it is
difficult to identify a source of uncertainty that could give rise to
systematic errors affecting the result by more than $30\%$.  Even with such an
uncertainty, the possibility to have all DM in the form of NRP sterile
neutrinos is ruled out by the comparison of \lya results with X-ray upper
bounds~\cite{Boyarsky:08c}.

In the RP case, \lya bounds have not been derived yet. However, in
Ref.~\cite{Boyarsky:08c}, a $\Lambda$CWDM model -- containing a
mixture of WDM (in the form e.g. of NRP sterile neutrinos) and Cold
Dark Matter (CDM) -- was analyzed. Below we will show that although
the phase-space distribution of RP sterile neutrinos does not coincide
exactly with such mixed models, some results can be inferred from the
$\Lambda$CWDM analysis. In particular, we will show that for each mass
$\ge 2\kev$, there is at least one value of the lepton asymmetry for
which the RP sterile neutrino model is fully consistent with \lya and
other cosmological data (this value of the lepton asymmetry is natural
within the \numsm).

\begin{figure}
  \centering
  \includegraphics[width=\linewidth]{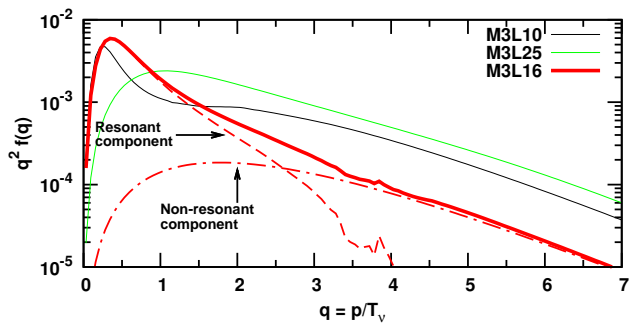}
  \caption{Characteristic form of the RP sterile neutrino distribution
    function for $m_s^\rp =3\kev$ and various values of the lepton
    asymmetry parameter $L_6$. The spectrum for $L_6=16$ (red solid line)
    is shown together with its resonant (dashed) and non-resonant
    (dashed-dotted) components. All these spectra have the same shape
    for $q\gtrsim 3$.}
  \label{fig:rp-spectra}
\end{figure}

\noindent\textbf{Spectra of RP sterile neutrino.}  DM production in the RP
scenario occurs in two stages~\cite{Shi:98,Laine:08a}.  In presence of a
lepton asymmetry, the conditions for resonant oscillations (related to the
intersection of dispersion curves for active and sterile neutrinos) are
fulfilled for temperatures of few hundred MeV. Later, at $T\sim 150 \,
(m_s^\rp/\mathrm{keV})^{1/3}\mev$, non-resonant production takes place.  As a
result, the primordial velocity distribution of sterile neutrinos contains a
narrow resonant (\emph{cold}) component and a non-resonant (\emph{warm}) one.
Its exact form can be computed by taking into account the expansion of the
Universe, the interaction of neutrinos with the content of the primeval plasma
and the changes in the lepton asymmetry resulting from DM production. 
In this work, we used the spectra
computed in Ref.~\cite{Laine:08a}.

Characteristic forms of the spectra are shown in Fig.~\ref{fig:rp-spectra}.
They are expressed as a function of the comoving momentum $q \equiv p/T_\nu$
($T_\nu$ being the temperature of active neutrinos), and depend on the lepton
asymmetry parameter~\cite{Shaposhnikov:08a} $L_6 \equiv 10^6 (n_{\nu_e} -
n_{\bar\nu_e})/s$ ($s$ being the entropy density). In rest of this work, the
notation M2L25 would refer to a model with $m_s^\rp=2\kev$ and $L_6=25$.  The
shape of the non-resonant distribution tail depends on the mass, but not on
$L_6$. For $q\gtrsim 3$, the distribution is identical to a rescaled NRP
spectrum~\cite{Asaka:06c} with the same mass (red dashed-dotted line on
Fig.~\ref{fig:rp-spectra}). We call this rescaling coefficient the \emph{warm
  component fraction} $\fnrp$. For the few examples shown in
Fig.~\ref{fig:rp-spectra}, the M3L16 models corresponds to $\fnrp \simeq
0.12$, M3L10 to $\fnrp \simeq 0.53$ and M3L25 to $\fnrp \simeq 0.60$.  The
maximum of $q^2 f(q)$ for the NRP component occurs around $q\approx 1.5-2$.
We define the \emph{cold component} to be the remaining contribution: its
distribution is given by the difference between the full spectrum and the
rescaled NRP one (red dashed line on Fig.~\ref{fig:rp-spectra}), and peaks
around $q_\text{res}\sim 0.25-1$. Its width and height depend on $L_6$ and
$m_s^\rp$.

The DM clustering properties can be characterized qualitatively by the
particle's \emph{free-streaming horizon} (see e.g.~\cite{Boyarsky:08c}
for definition), proportional to its average velocity $\langle
q\rangle/m$.  In the RP case, the dependence of
the average momentum $\langle q\rangle$ on $m_s^\rp$ and $L_6$ is not
monotonic. For a given mass, the RP model departing most from
an NRP model is the one with the smallest
$\langle q\rangle$ (i.e., with the most significant cold component).
For each mass, there is indeed a value of $L_6$ minimizing $\langle
q\rangle$, such that $\langle q\rangle_{min} \approx 0.3 \langle
q\rangle_\nrp$ (cf.~\cite{Laine:08a}). This minimum corresponds to
lepton asymmetries that are likely to be generated within the \numsm.
For instances, for $m_s^\rp=2$, 3 or 4~keV,
the spectra with the smallest average momentum are M2L25, M3L16 and M4L12,
all having $\fnrp \lesssim 0.2$.

For a quantitative analysis, we computed the power spectrum of matter density
perturbations $P_\rp(k,z)$ for these models. The standard software (i.e.
\textsc{camb}~\cite{Lewis:99}) is not immediately appropriate for this
purpose, as it only treats massive neutrinos with a Fermi-Dirac primordial
distribution. To adapt it to the problem at hand, we modified \textsc{camb} so
that it could take arbitrary spectra as input data files.  We analyzed the
spacing in momentum space needed in order to obtain precise enough results,
and implemented explicit computations of distribution momenta in
\textsc{camb}.  We cross-checked our results by modifying another linear
Boltzmann solver -- \textsc{cmbfast}~\cite{Seljak:96}, 
implementing a treatment of
massive neutrinos with arbitrary \emph{analytic} distribution function.

\begin{figure}[t]
  \centering
  \includegraphics[width=\linewidth]{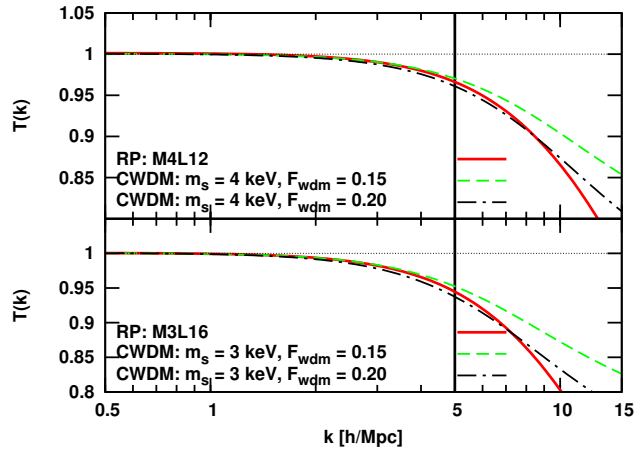}
  \caption{TFs for the models M4L12 (top) and M3L16 (bottom), together with
    CWDM spectra for the same mass and $F_{\wdm} \simeq 0.15$ or 0.2.}
  \label{fig:rp4_vs_cwdm}
\end{figure}

\begin{figure}
  \centering
  \includegraphics[width=\linewidth]{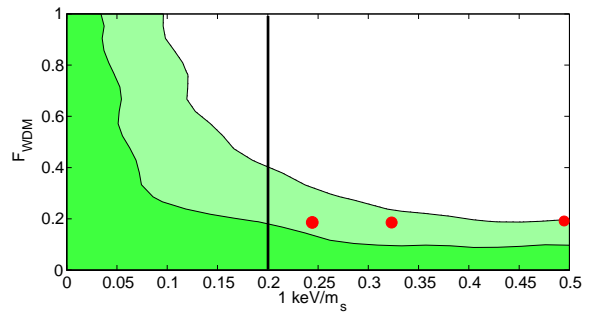}
  \caption{1 and 2-$\sigma$ bounds from WMAP5 and SDSS \lya data on
    $\Lambda$CWDM parameters. Red points correspond to
    approximations for the models
    M4L12, M3L16, M2L25.  Results to the left of black vertical lines
    were already reported in~\cite{Boyarsky:08c}.}
  \label{fig:cwdm}
\end{figure}

To separate the influence of primordial velocities on the evolution of
density perturbations from that of cosmological parameters, it is
convenient to introduce the transfer function (TF) $T(k) \equiv
\bigl[P_\rp(k)/P_{\lcdm}(k)\bigr]^{1/2}$. Figure~\ref{fig:rp4_vs_cwdm}
shows the transfer function of the models M3L16 and M4L12.  The TF
becomes smaller than one above the wave number associated with the
free-streaming horizon today, $k_\textsc{fsh} \approx 0.5 \,
({m_s^\rp}/\mathrm{1\kev}) \, h \!  \mpc^{-1}$
(c.f.~\cite{Boyarsky:08c}).  We see that for a large range of $k$
values above $k_\textsc{fsh}$, roughly $k \lesssim 5 \,
k_\textsc{fsh}$, the transfer function $T_\rp(k)$ is very close to
$T_\lcwdm(k)$ for the same mass and warm component fraction $\Fwdm
=\fnrp$.  On smaller scales, $T_\rp(k)$ decreases faster, since the
cold component of RP sterile neutrinos also has a non-negligible
free-streaming scale. For all values of the mass studied in this work,
$m_s^\rp \geq 2$~keV, the discrepancy appears above 5$~h/\! \mpc$
(vertical line in Fig.~\ref{fig:rp4_vs_cwdm}), i.e.  above the maximum
scale in the three-dimensional power spectrum to which current \lya
data are sensitive. Hence, for the purpose of constraining RP sterile
neutrinos with \lya data, it is possible to use the results obtained
in the $\lcwdm$ case.
In Ref.~\cite{Boyarsky:08c}, we presented the results of a WMAP5 plus SDSS
\lya data analysis for $\Lambda$CWDM models with $m_s^\rp \geq 5$~keV. In
Fig.\ref{fig:cwdm}, we show the Bayesian credible region for the mass and the
warm component fraction, now extended till $m_s^\nrp =
2$~keV~\footnote{Thermal velocities of NRP neutrinos may become important
  below $5\kev$~\cite{Boyarsky:08c}. To be conservative, this mass region was
  excluded from the results of~\cite{Boyarsky:08c}, but we put it back here.
  We see that 1 and $2\sigma$ contours become nearly horizontal for $m\lesssim
  5\kev$ (c.f.  Fig.~\ref{fig:cwdm}). We do not expect thermal velocities to
  affect this conclusion. For detailed bounds on RP sterile neutrinos, we plan
  to explore the precise dependence of N-body simulation results on thermal
  velocities elsewhere.}.

Fig.~\ref{fig:rp4_vs_cwdm} demonstrates that for the models M3L16 and M4L12,
the function $T_\rp(k)$ lies above $T_\lcwdm(k)$ for the same masses and
$\Fwdm = 0.2$ (at least, in the range of wave numbers probed by Ly-$\alpha$
data): so, it can only be in better agreement with cosmological data. We
checked that the same is true for M2L25.
%
However, $\Lambda$CWDM models with $m_s^\rp= 2, 3, 4 \, {\rm keV}$ and $\Fwdm
= 0.2$ are within the 2-$\sigma$ contour of Fig.~\ref{fig:cwdm}.  We conclude
that M2L25, M3L16 and M4L12 are clearly allowed by the data. For
larger mass (and still minimal $\langle q \rangle$), the free-streaming
horizon is smaller, and agreement with observations can only become easier.
Therefore, we see that for each mass $m_\rp \gtrsim 2\kev$ there exists
\emph{at least} one value of the lepton asymmetry for which RP sterile
neutrinos are perfectly compatible with WMAP5 and SDSS \lya data.

\begin{figure}
  \centering
  \includegraphics[width=1.\linewidth]{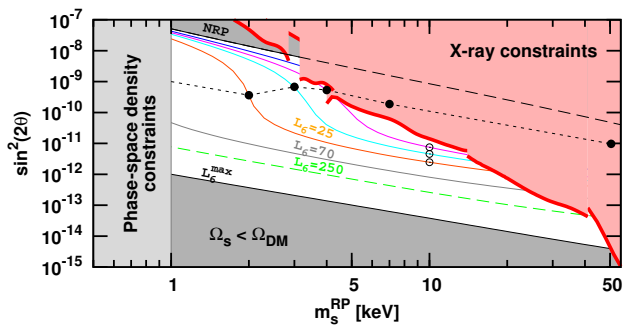}
  \caption{Region of masses and mixing angles for RP sterile neutrinos
    consistent with existing constraints.}
  \label{fig:exclusions}
\end{figure}

Fig.~\ref{fig:exclusions} shows the range of masses and mixing angles
consistent with constraints from phase-space
density~\cite{Boyarsky:08a} (left shaded region), from
X-rays~\cite{Boyarsky:06c,Boyarsky:06d} (upper right corner, shaded in
red) and providing the correct DM abundance (curves
between the lines ``NRP'' and $L_6^{max}$: from top to bottom $L_6=8,
12, 16, 25, 70, 250$). The black dashed line shows approximately the
RP models with minimal $\langle q\rangle$ for each mass, i.e., the
family of models with the largest cold component.  We have seen that
all black filled circles along this line and with $m_s^\rp \geq 2\kev$
are compatible with \lya bounds. In addition, those with $m_s^\rp \leq
4\kev$ are also compatible with X-ray bounds (this
conclusion does not change with the new results of
Ref.~\cite{Loewenstein:2008yi}). Note that above $4\kev$, \lya data allows
increasingly high WDM fractions, so that agreement with both \lya and
X-ray bounds can be maintained with larger values of $L_6$.  This is
very clear e.g. for the models M10L25, M10L16 and M10L12, allowed by
X-ray data (open circles on Fig.\ref{fig:exclusions}), and consistent
with \lya data since for $m_s^\rp = 10\kev$, up to 100\% of WDM is
allowed at the 2-$\sigma$ level (c.f.  Fig.\ref{fig:cwdm}).



\emph{In conclusion,} we showed in this work that sterile neutrino DM
with mass $\ge 2\kev$ is consistent with all existing constraints. A
sterile neutrino with mass $\sim 2\kev$ is an interesting WDM
candidate, as it may affect structure formation on galactic
scales. This range of masses and corresponding mixing angles is
important for laboratory and astrophysical searches.

To determine the precise shape of the allowed parameter range (which may
continue below 2~keV, see Fig.~\ref{fig:exclusions}), one should perform
specific hydrodynamical simulations in order to compute the flux power
spectrum on a grid of ($m_s^\rp$, $L_6$) values, and compare with \lya data.
We leave this for future work.

The \numsm does not require new particles apart from the three sterile
neutrinos.
Extensions of this model may include a scalar field providing Majorana masses
to sterile neutrinos via Yukawa couplings~\cite{Shaposhnikov:06,Kusenko:06a}.
Then, sterile neutrino DM can also be produced by the decay of this scalar
field, and also contain a cold and a warm component.
We expect a similar range of masses and mixing angles to be
allowed by \lya data.  The quantitative analysis of this model is also left
for future work.


\textbf{Acknowledgments.} We thank A.~Kusenko, M.~Laine, M.~Shaposhnikov,
S.~Sibiryakov for useful discussions.  J.L. acknowledges support from the EU
network ``UniverseNet'' (MRTN-CT-2006-035863). O.R.  was supported by the
Swiss Science Foundation.

\bibliographystyle{apsrev}%

\let\jnlstyle=\rm\def\jref#1{{\jnlstyle#1}}\def\aj{\jref{AJ}}
  \def\araa{\jref{ARA\&A}} \def\apj{\jref{ApJ}\ } \def\apjl{\jref{ApJ}}
  \def\apjs{\jref{ApJS}} \def\ao{\jref{Appl.~Opt.}} \def\apss{\jref{Ap\&SS}}
  \def\aap{\jref{A\&A}} \def\aapr{\jref{A\&A~Rev.}} \def\aaps{\jref{A\&AS}}
  \def\azh{\jref{AZh}} \def\baas{\jref{BAAS}} \def\jrasc{\jref{JRASC}}
  \def\memras{\jref{MmRAS}} \def\mnras{\jref{MNRAS}}
  \def\pra{\jref{Phys.~Rev.~A}} \def\prb{\jref{Phys.~Rev.~B}}
  \def\prc{\jref{Phys.~Rev.~C}} \def\prd{\jref{Phys.~Rev.~D}}
  \def\pre{\jref{Phys.~Rev.~E}} \def\prl{\jref{Phys.~Rev.~Lett.}}
  \def\pasp{\jref{PASP}} \def\pasj{\jref{PASJ}} \def\qjras{\jref{QJRAS}}
  \def\skytel{\jref{S\&T}} \def\solphys{\jref{Sol.~Phys.}}
  \def\sovast{\jref{Soviet~Ast.}} \def\ssr{\jref{Space~Sci.~Rev.}}
  \def\zap{\jref{ZAp}} \def\nat{\jref{Nature}} \def\iaucirc{\jref{IAU~Circ.}}
  \def\aplett{\jref{Astrophys.~Lett.}}
  \def\apspr{\jref{Astrophys.~Space~Phys.~Res.}}
  \def\bain{\jref{Bull.~Astron.~Inst.~Netherlands}}
  \def\fcp{\jref{Fund.~Cosmic~Phys.}} \def\gca{\jref{Geochim.~Cosmochim.~Acta}}
  \def\grl{\jref{Geophys.~Res.~Lett.}} \def\jcp{\jref{J.~Chem.~Phys.}}
  \def\jgr{\jref{J.~Geophys.~Res.}}
  \def\jqsrt{\jref{J.~Quant.~Spec.~Radiat.~Transf.}}
  \def\memsai{\jref{Mem.~Soc.~Astron.~Italiana}}
  \def\nphysa{\jref{Nucl.~Phys.~A}} \def\physrep{\jref{Phys.~Rep.}}
  \def\physscr{\jref{Phys.~Scr}} \def\planss{\jref{Planet.~Space~Sci.}}
  \def\procspie{\jref{Proc.~SPIE}} \let\astap=\aap \let\apjlett=\apjl
  \let\apjsupp=\apjs \let\applopt=\ao

\end{document}